# PWO crystals for CMS electromagnetic calorimeter: studies of the radiation damage kinetics

G.Drobychev[1*], E. Auffray[2], V.Dormenev[1], M.Korzhik[1], P.Lecoq[2], A. Lopatic[1], P.Nedelec[3], J.-P.Peigneux[3], D.Sillou[3]

[1]*Institute for Nuclear Problems, Minsk, 220050, Belarus*

[2] *CERN, CH-1211, Geneve, Switzerland*

[3]*LAPP, 74941, Annecy-le-Vieux, France*

## Abstract

Kinetics of radiation damage of the PWO crystals under irradiation and recovery were studied. Crystals were irradiated with dose corresponding to average one expected in the electromagnetic calorimeter (working dose irradiation). Radiation damage and recovery were monitored through measurements of PWO optical transmission. An approach is proposed which allows evaluating the influence of the PWO crystals properties on the statistical term in the energy resolution of the electromagnetic calorimeter. The analysis also gives important information about the nature of the radiation damage mechanism in scintillation crystals. The method was used during development of technology of the mass production of radiation hard crystals and during development of methods for crystals certification.

[*] Corresponding author. Tel.: +37517-2264221; fax: +37517-2265124; e-mail: Gleb.Drobychev@cern.ch.

## Introduction

The CMS (Compact Moon Solenoid) experimental setup is under construction now at the LHC accelerator facility in the European Laboratory for Particle Physics (CERN, Switzerland) [1]. The Electromagnetic calorimeter (ECAL) is one of the major elements of the setup. The ECAL contains a multiplicity of total absorption detectors based on PWO scintillation crystals, optically linked with avalanche photodiode (APD) used as photo detector. The gamma radiation, entering the detector, initiates an electromagnetic shower in the crystal. Electrons and positrons loose their energy by ionization in the bulk of the crystal which causes an emission of scintillation light. The scintillation photons collected on the photodetector generate an electric signal registered by the read-out electronics. It is designed to detect Higgs bosons in a range of masses of 90 GeV $\leq m_H$ $\leq$150 GeV through the $H\to\gamma\gamma$ decay channel. Because this Higgs bosons decay is a relatively rare event, long time measurements during several working cycles of the accelerator are required. A year cycle of the LHC accelerator includes three 60 days operation periods with full radiation charge and 14 days intervals in between, when there is no radiation. After these three cycles there is a complete stop of accelerator for about 150 days. High level irradiation appears in the ECAL barrel during LHC operation. Depending on the angle distribution of the shower, the maximum dose rate may be as high as 0.25 Gy/h.

The quantity of registered photons depends on the light collection efficiency, which, in turn, depends on the crystal transparency. As was shown in [2, 3] radiation-induced optical absorption in PWO crystals in wide region of radiation doses is not caused by damage of radiation centers but mainly is the result of color centers formation at the point structure defects. Formation of color centers induces a decrease in the PWO optical transparency which in turn causes a worsening of the ECAL energy resolution [4]. This is the reason of the importance of the study of radiation damage and recovery processes of PWO scintillation crystals.

This paper presents results of research on the PWO optical transmission kinetics under irradiation which were carried out at the General Irradiation Facility (GIF) at CERN ($Cs^{137}$ γ-source, $E_\gamma$~0.66 MeV). Dose and dose rate correspond to the average values expected in the ECAL during operation periods (dose rate 0.15 Gy/h, absorption dose 6 Gy). The experimental setup provided a continuous monitoring of the PWO optical transmission. Results of the crystals transmission measurements were compared with the results of PWO light yield measurements performed with high energy electrons beam (100 GeV). A detailed analysis was performed from data of three crystals.

## Analysis of kinetics of the PWO optical transmission damage and recovery

Fig. 1 presents spectra of luminescence and radiation induced absorption coefficient of optical transmission of typical PWO crystal. We studied relaxation kinetics of the induced absorption at the 420 nm wavelength of doped and non-doped crystals, because luminescence maximum is positioned in this region [3].

Radiation induced absorption is determined through the following equation:

$$\ddot{A}k = \frac{1}{L}\cdot \ln\left[T_1/T_2\right] \qquad (1),$$

with $L$ – the crystal length, $T_1$ and $T_2$ – the optical transmissions of crystal before and after irradiation.

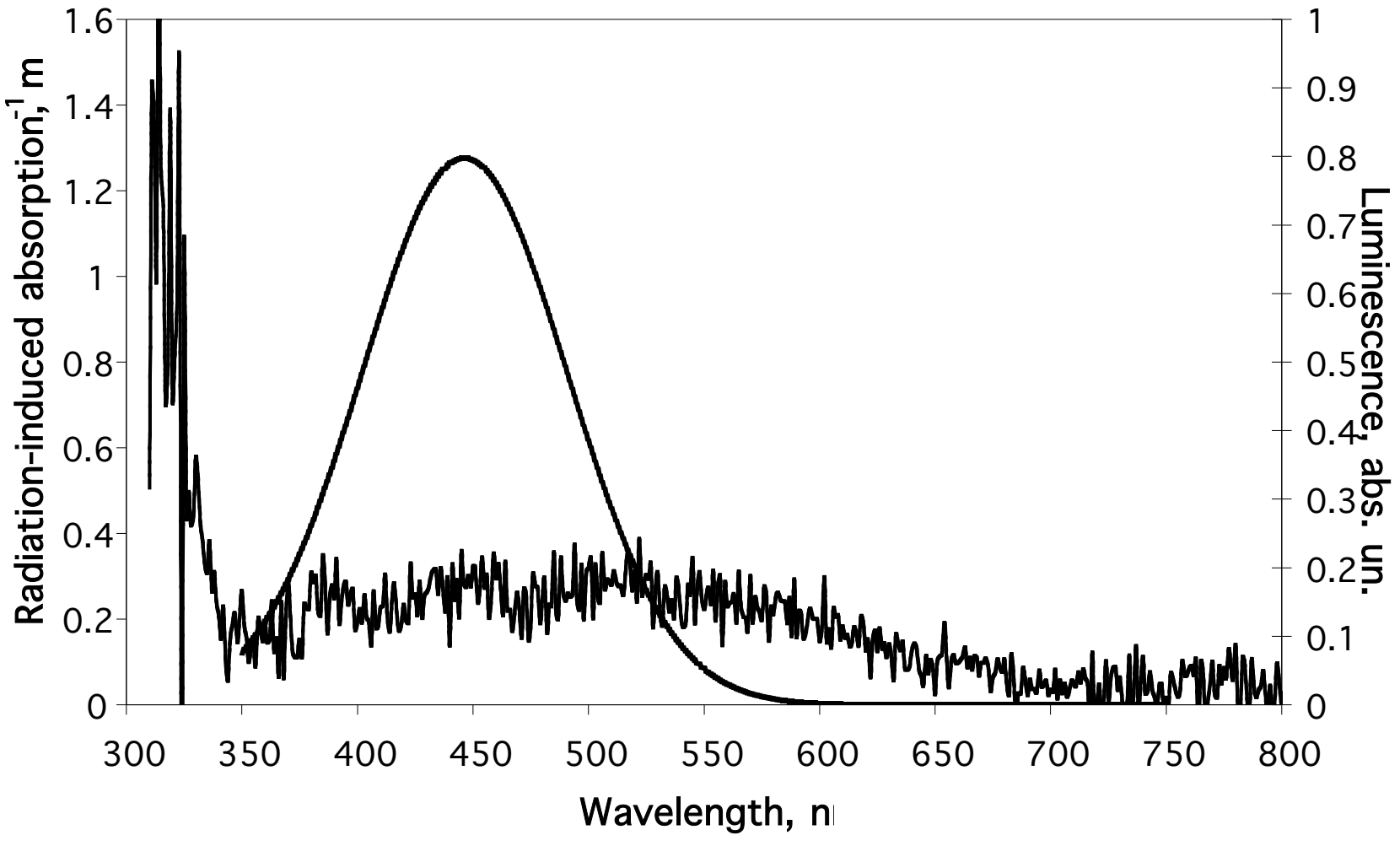

**Fig. 1: Spectra of luminescence and radiation induced coefficient of optical transmission of typical PWO crystal.**

As the analysis shows, at working dose irradiation, an evolution of the optical transmission at the 420 nm wavelength of all samples is well fitted by the following equation (see Fig. 2):

$$\delta T(t) = 1 - \Sigma_i \alpha_i \cdot \exp(-t/\tau_i) \qquad (2),$$

with $\alpha_I$ – contribution to the $\delta T$ during irradiation period and $\tau_I$ – time constant of the particular color center formation.

The exponential approximation corresponds to the defects decaying through thermo activation only. The contribution of a large amount of components can be well approximated under the above conditions of irradiation: time discreteness, accuracy of δT measurements and relative contribution of $\alpha_i$ by two exponentials with time constants $10^3$ s and $10^5$ – $10^6$ s. The fast component resulted from the thermo activation of the $(WO_3\text{–}WO_3)^{2-}$ centers. However, the slow component cannot be linked to the thermo induced recombination. It is known that electron and hole centres formed on PWO point structure defects have thermo activation energies in the region from 0.03 to 0.7 eV and the longest activation time of these centers is about $10^3$ s. That is why, beyond thermo-activation of centers and the following recombination of formed carriers, other recombination processes, such as, those caused by electrons tunneling, should exist [5,6]. This tunneling effect means that electron due to its quantum mechanical nature can be localized with some probability at rather long distance from electronic center. There is some probability that an electron will appear in the vicinity of a hole center located rather distant (~ 100 Å) from the electronic center and that it recombines with this hole center.

That is why we consider a recovery of optical transmission as a combination of two processes: decay of color centers through thermo activation for the fast component and diffusion-tunnel recombination for slow component. The recovery curve taking into consideration a diffusion –tunnel recombination mechanism should follow the following equation [7]:

$$\frac{n(t)}{n(0)} = \sum_i \frac{n_i(0)}{n(0)} e^{-\frac{t}{\tau_i}} + \frac{n_t(0)}{n(0)} \times \frac{1}{[1 + n_t(0)(\pi a^3/6)\ln^3 \nu t]} \qquad (3)$$

where $a = h/2\pi[2m(U_{\max} - E)]^{1/2}$, $n(0)$ – starting concentration of the electronic defects, $m$ –electron mass, $(U_{max} - E)$ difference between potential barrier and center's energy, ν – frequency of electron oscillation in the oscillation potential, which is well approximated by the average frequency of phonon spectrum of crystal (500 cm$^{-1}$)[8], $n_i(0)$ – contribution of the *i–th* exponential component into the induced absorption at $t$ = 0, and $n_t(0)$ – contribution of tunnel recombination centers into induced absorption at $t$ = 0.

The results of experimental data fits according to equations (2,3) are presented in Fig. 2 and in Tables 1, 2. As one can see the experimentally observed recombination of optical transmission is well approximated by a linear combination of the same number of exponents than was used for the damage fit. Fractional contributions of components at damage and recovery are approximately corresponding to each other. That is why for further calculations we used a more simple two exponents approximation instead of more complex calculations according to equation (3).

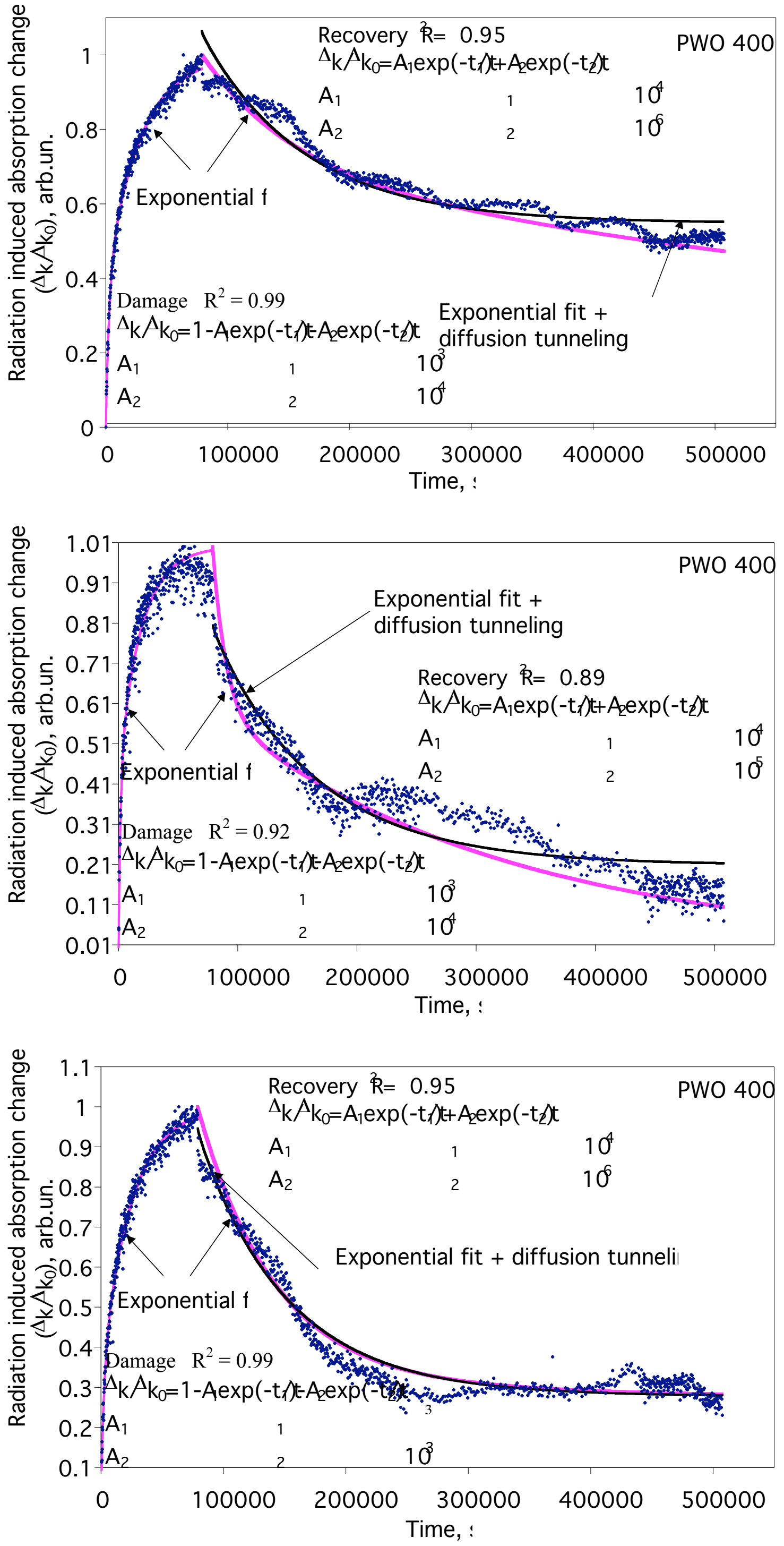

**Fig. 2 a–c. Kinetics of damage and recovery of PWO No. 4002 (a), 4004 (b), 4005 (c); dots – experimental data, lines – two exponential fit of damage and recovery, dashed lines fit with diffusion – tunnel model.**

**Table 1. Results of diffusion – tunnel approximation.**

| Sample | _$_i$, s | $\Delta k_{exp}(0)$, cm$^{-1}$ | $n(0)$, cm$^{-3}$ | $\Delta k_{tun}(0)$, cm$^{-1}$ |
|---|---|---|---|---|
| 4002 | $8.5\times10^4$ | 0.084 | $1.7\times10^{17}$ | 0.101 |
| 4004 | $8.6\times10^4$ | 0.082 | $6.7\times10^{17}$ | 0.041 |
| 4005 | $7.3\times10^4$ | 0.092 | $5.6\times10^{17}$ | 0.052 |

The chosen simplified model allowed to carry out evaluations of crystals optical transmission change during LHC accelerator working cycle, which includes active phase (transmission of defects to the color centers, i.e. radiation damage) lasting $t_d$ accepted as 6 hours, and recharge phase (color centers decay, i.e. recovery of crystal optical transmission) lasting $t_r$ accepted as 2 hours. We also accepted that dose rate during tests corresponds to the average dose rate during the active phase [1].

According to [9], under the exponential approximation and in stationary regime, in which optical transmission change during active phase and recharge phase are equal in absolute value we can write the following equation for relative change of defects number during $t_d$ and relative change of color centers number during $t_r$:

$$\lambda_d = -\frac{dN_d}{dt \times N_d(t)} \qquad (4),$$

where $\lambda_d$ is the speed of the defect transformation to the color center (damage), which we consider to be constant and proportional to the dose rate, and $N_d(t)$ is the defect number at time $t$ after the start of the irradiation;

$$\lambda_r = \frac{dN_r}{dt \times N_r(t)} \qquad (5),$$

where $\lambda_r$ is the speed of color center transformation to the defect (recovery), which we consider to be constant, and $N_r(t)$ is the color center number at time $t$ after the start of irradiation.

The following equation gives the correspondence between the number of defects and of color centers:

$N_d + N_r = N_0$ (6),

where $N_0$ – number of defects before start of irradiation.

Differential equation of kinetics of defects number:

dN$_d$/dt = – _$_d$N$_d$ + _$_r$(N$_0$ – N$_d$) (7).

General solution of kinetic equation (7) for the defects number (8) and color centers (9):

N$_d$ = (N$_0$ /(_$_d$ + _$_r$)) (_$_r$ + __ exp (– (_$_d$ + _$_r$)t)) (8),

N$_r$ = (N$_0$ /(_$_d$ + _$_r$)) (_$_d$ – __ exp (– (_$_d$ + _$_r$)t)) (9),

where __ is a constant which depends on the boundary conditions.

Time constant of coming to stationary regime is:

_ = 1/(_$_d$ + _$_r$) (10).

Defects recovery after stop of irradiation is described by the following differential equation:

dN$_d$/dt = _$_r$(N$_0$ – N$_d$) (11)

Then, there is a general solution for the kinetics equation (11) for number of defects (12) and number of color centers (13):

N$_d$ = N$_0$ (1 – __ exp (– _$_r$t)) (12),

N$_r$ = N$_0$ __ exp (– _$_r$t) (13),

where __ is a constant which depends on boundary conditions.

Experimental data were fitted with exponential functions according to equations (8), (9) (during irradiation) and equations (12), (13) (after irradiations).

Under stationary regime of PWO transmission change at LHC operation a number of color centers formed during time $t_d$ is $-\Delta N_d(t_d)$ and corresponds to the quantity of defects $\Delta N_r(t_r)$ recovered during the recharge time $t_r$:

$$-\Delta N_d(t_d) = \Delta N_r(t_r) \quad (14)$$

Let's introduce variables:

$$f_r = \exp(-\lambda_r t_r) \quad (15),$$

$$f_d = \exp(-(\lambda_d + \lambda_r)t_d)) \quad (16),$$

From (8), (9) and (12), (13) with condition (14), a following equation can be written for relative quantity of defects at the beginning and after completing of recovery cycle $\alpha_r$ and $\alpha_d$ correspondingly.

For $\alpha_r$

$$\alpha_r = (1 - f_r (f_d - (\lambda_d/(\lambda_d + \lambda_r)) (f_d - 1)))/(1 - f_r f_d) \quad (17).$$

For $\alpha_d$

$$\alpha_d = (1 - f_r f_d + (\lambda_d/(\lambda_d + \lambda_r))( f_d - 1))/(1 - f_r f_d) \quad (18).$$

Then, following equation can be written for a relative change of transmission at stationary regime at the beginning and after completing of recovery cycle $\delta T_r$ and $\delta T_d$ correspondingly:

$$\delta T_r = 1 - \alpha_r = ((\lambda_d/(\lambda_d + \lambda_r)) f_r (1 - f_d))/(1 - f_r f_d) \quad (19),$$

and

$$\delta T_d = 1 - \alpha_d = (\lambda_d/(\lambda_d + \lambda_r)) ((1 - f_d))/(1 - f_r f_d) \quad (20).$$

During irradiation cycle a relative change of transmission $\delta T_$, taking into consideration (19), (20), is following:

$$\delta T_ = \delta T_d - \delta T_r = ((\lambda_d/(\lambda_d + \lambda_r))(1 - f_r)(1 - f_d))/(1 - f_r f_d) \quad (21).$$

A characteristic time of coming to the stationary regime according the criteria of two time constant is:

$$\tau = 2/(\lambda_d + \lambda_r) \quad (22).$$

A correspondence between time constants at irradiation and at recovery, $\tau_d$ and $\tau_r$, and corresponding speeds of formation and decay of color centers are determined by the following equations:

$$\lambda_r = 1/\tau_r \quad (23)$$

and

$$\lambda_d = 1/\tau_d - 1/\tau_r \quad (24).$$

The results of experimental data analysis of damage and recovery time constants $\tau_d$, $\tau_r$,, amplitudes $A_d$, $A_r$ and contributions to the optical transmission change and results of calculations of $\lambda_r$, $\lambda_d$, $\delta T_r$, $\delta T_d$ are listed in Table 2. Choice of correspondence of the $\tau_d$ and $\tau_r$ of the same i–th group of centers (i = 1, 2) was made taking into account that the time constant at irradiation is always smaller than the one at recovery and that the contributions of the same group of centers to the optical transmission change $\delta T$ must correspond to each other within the accuracy of the approximation.

As one can see from the obtained data (see table. 2 and Fig. 3), in stationary regime after recharge, a relative change of defects quantity for all components is from 50 to 100 % and after simulation of the active cycle a quantity of color centers becomes close to saturation for the dose rate and irradiation schedule ($t_d$ = 6 hours and $t_r$ = 2 hours) used.

From experimental data we can observe a presence in damage and recovery of components with time constant of several hundreds seconds. Relative contribution of these components into $\delta T$ is about 10–15 %. Because experimental data were obtained with a measurements sampling period of 240 s, it is impossible to define precisely the "fast" components parameters without an increase of the dose rate and of the measurements frequency.

**Table 2. Results of experimental data analysis on the radiation damage and recovery of PWO crystals at working dose and results of calculations of the damage and recovery characteristics in conditions of LHC cycle.**

| Parameter | | Experimental data analysis | | | | | Calculations | | | |
|---|---|---|---|---|---|---|---|---|---|---|
| Number | | $\tau_d$, s | $A_d$ | $\tau_r$, s | $A_r$ | $\delta T$, % | $\lambda_d$, $s^{-1}$ | $\lambda_r$, $s^{-1}$ | $\delta T_r$, % | $\delta T_d$, % |
| 4002 | 1 | $2.7\times10^3$ | 0.38 | $7.5\times10^4$ | 0.31 | 35 | $3.6\times10^{-4}$ | $1.3\times10^{-5}$ | 91 | 8.7 |
| | 2 | $2.8\times10^4$ | 0.62 | $1.1\times10^6$ | 0.69 | 62 | $3.5\times10^{-5}$ | $0.9\times10^{-6}$ | 98 | 0.9 |
| 4004 | 1 | $1.7\times10^3$ | 0.44 | $1.1\times10^4$ | 0.41 | 43 | $5.0\times10^{-4}$ | $0.9\times10^{-4}$ | 52 | 44 |
| | 2 | $1.9\times10^4$ | 0.56 | $2.5\times10^5$ | 0.59 | 57 | $4.8\times10^{-5}$ | $5.0\times10^{-6}$ | 94 | 2.5 |
| 4005 | 1 | $2.2\times10^3$ | 0.32 | $5.5\times10^6$ | 0.30 | 31 | $4.5\times10^{-4}$ | $1.8\times10^{-7}$ | 100 | 0 |
| | 2 | $2.4\times10^4$ | 0.68 | $6.4\times10^4$ | 0.70 | 69 | $2.7\times10^{-5}$ | $1.5\times10^{-5}$ | 74 | 10 |

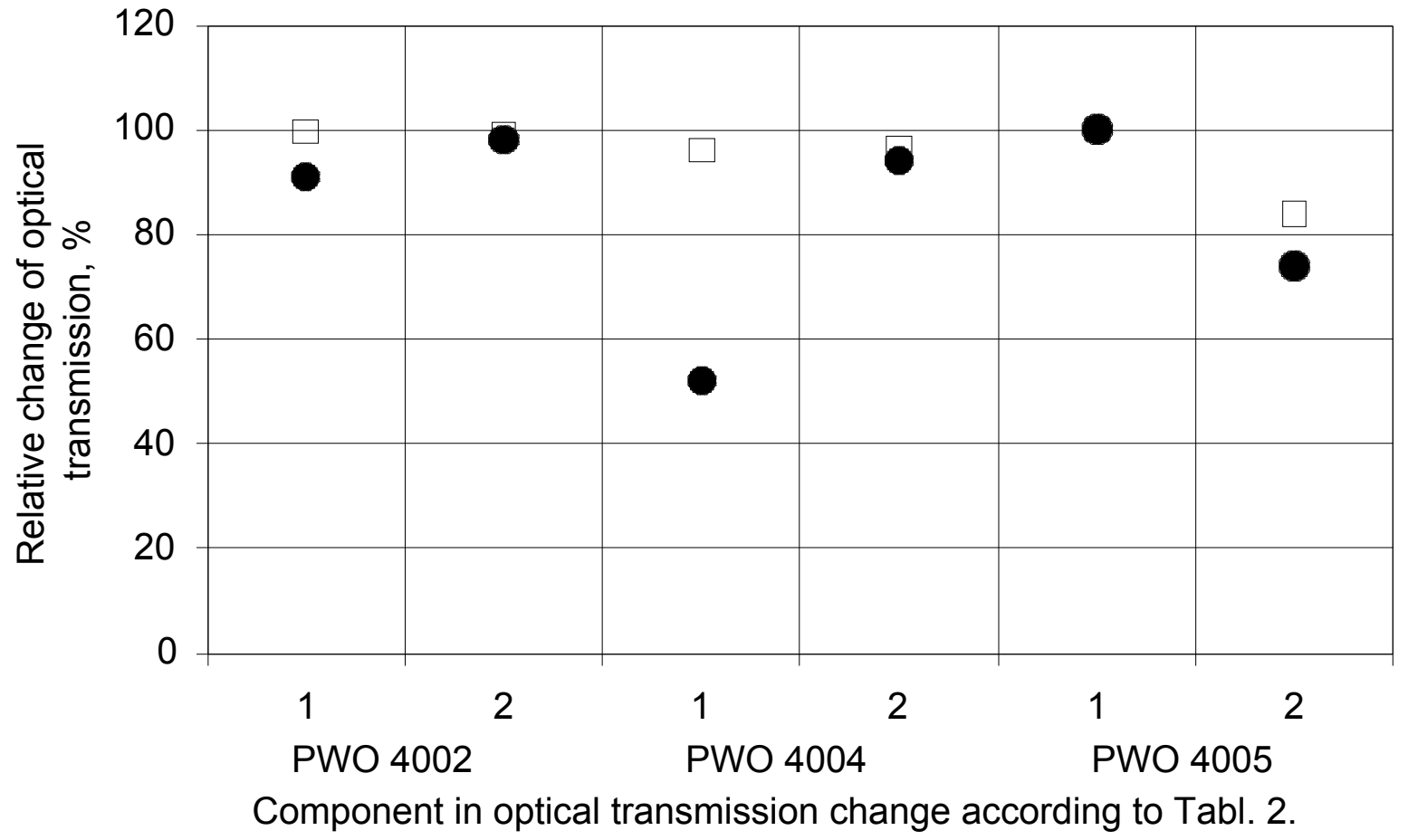

**Fig. 3. Contributions of defects in slow and fast components in PWO radiation damage and recovery; circles – relative change of transmission at the beginning of the recovery cycle ($\delta T_r$), squares – relative change of transmission at the end of the recovery cycle ($\delta T_d$).**

Taking into account the obtained data, the time necessary to reach the stationary regime is about 3 working cycles of LHC, i.e. about 24 hours.

From Table 2 one can see that the maximal time of PWO recovery after irradiation is $5.5\times10^6$ s or 64 days. It means that a period of complete stop of accelerator is sufficient for the complete recovery of crystals. At the same time, one can see that during 14–days period of accelerator stop, a part of the crystals will not completely recover. As one can see during the 14 days period one can expect a recovery till the level of 80 % from original value.

To extend the results of the optical transmission kinetic measurements on the technological lot of PWO, a test of PWO sampling consisting of about 1000 crystals with various dose rates corresponding to different zones of the ECAL have to be carried out.

## Conclusion

Kinetics of radiation damage of the PWO crystals under irradiation and recovery were studied. An approach to evaluate an influence of individual PWO characteristics in strong radiation fields on the photostatistic term in ECAL energy resolution is proposed.

Kinetics of radiation damage and recovery of several PWO crystals is analyzed. The analysis results show the presence of slow components which cannot have thermo activation origin. To explain a presence of these components a diffusion tunnel process of deep traps decay is proposed.

The collected data shows that a more detailed analysis of crystals behavior under irradiation requires the continuation of the regular research on the PWO radiation damage and recovery kinetics.

The proposed approach was used during development of the mass production technology of radiation hard crystals and during development of methods for crystals certification.

**Acknowledgments.**

This work was done in frame of the Project 1718P of the International Scientific and Technical Center. The research was also supported in part by the State program for basic research carried out in the Belarus State University.

Authors would like to thank Dr. Ph. Bloch and T. Camporezi from CERN and A.N. Annenkov from BTCP for interest to this work and valuable discussions. Authors are very grateful to Dr.A.Singovsky, Dr. P.Rebecchi, Dr. M. Ridel and Mr. V.Poireau from CERN for help in setting up and data taking, and to Dr. O. Missevitch, A. Fyodorov, V. Moroz†, A. Borisevich from INP for valuable discussions.

**References.**